\documentclass[aps,pra,twocolumn,
	amsmath,amssymb,amsfonts,floatfix,
	longbibliography,superscriptaddress,nobibnotes]{revtex4-2}
    
\usepackage[utf8]{inputenc}
\usepackage[T1]{fontenc} 
\usepackage{graphicx} 
\usepackage{bm}
\usepackage{placeins}

\usepackage[hypertexnames=false,colorlinks=true,citecolor=blue,
			linkcolor=blue,urlcolor=blue]{hyperref}

\usepackage{xfrac}

\usepackage{hyperref}

\usepackage[dvipsnames]{xcolor}
\definecolor{mygreen}{rgb}{0.0, 0.6, 0.0}
\definecolor{pjorange}{rgb}{0.8, 0.3, 0.0}
\definecolor{jlblue}{rgb}{0.2, 0.5, 0.7}

\graphicspath{{fig/}{./fig/}{.}}

\begin{document}

\title{Dimensional control of the band-gap crossover in layered lead iodide}

\author{M. Rosmus}
\email[e-mail: ]{marcin.rosmus@universite-paris-saclay.fr}
\affiliation{\mbox{Universit\'{e} Paris-Saclay, CNRS, Institut des Sciences Mol\'{e}culaires d’Orsay, 91405, Orsay, France}}

\author{A. Antezak}
\affiliation{\mbox{Universit\'{e} Paris-Saclay, CNRS, Institut des Sciences Mol\'{e}culaires d’Orsay, 91405, Orsay, France}}

\author{A. Ptok}
\affiliation{\mbox{Institute of Nuclear Physics, Polish Academy of Sciences, 
W. E. Radzikowskiego 152, PL-31342 Krak\'{o}w, Poland}}

\author{F. Fortuna}
\affiliation{\mbox{Universit\'{e} Paris-Saclay, CNRS, Institut des Sciences Mol\'{e}culaires d’Orsay, 91405, Orsay, France}}

\author{C. P. Sonny Tsotezem}
\affiliation{\mbox{Universit\'{e} Paris-Saclay, CNRS, Institut des Sciences Mol\'{e}culaires d’Orsay, 91405, Orsay, France}}

\author{E. M. Staicu Casagrande}
\affiliation{\mbox{Universit\'{e} Paris-Saclay, CNRS, Institut des Sciences Mol\'{e}culaires d’Orsay, 91405, Orsay, France}}

\author{A. Momeni}
\affiliation{\mbox{Universit\'{e} Paris-Saclay, CNRS, Institut des Sciences Mol\'{e}culaires d’Orsay, 91405, Orsay, France}}

\author{A. Ouvrard}
\affiliation{\mbox{Universit\'{e} Paris-Saclay, CNRS, Institut des Sciences Mol\'{e}culaires d’Orsay, 91405, Orsay, France}}

\author{C. Bigi}
\affiliation{\mbox{SOLEIL Synchrotron, L'Orme des Merisiers, D\'{e}partementale 128, 91190, Saint-Aubin, France}}

\author{M. Zonno}
\affiliation{\mbox{SOLEIL Synchrotron, L'Orme des Merisiers, D\'{e}partementale 128, 91190, Saint-Aubin, France}}

\author{A. Ouerghi}
\affiliation{\mbox{Universit\'{e} Paris-Saclay, CNRS,  Centre de Nanosciences et de Nanotechnologies, 91120, Palaiseau, France}}

\author{H. Khemliche}
\affiliation{\mbox{Universit\'{e} Paris-Saclay, CNRS, Institut des Sciences Mol\'{e}culaires d’Orsay, 91405, Orsay, France}}

\author{A. F. Santander-Syro}
\email[e-mail: ]{andres.santander-syro@universite-paris-saclay.fr}
\affiliation{\mbox{Universit\'{e} Paris-Saclay, CNRS, Institut des Sciences Mol\'{e}culaires d’Orsay, 91405, Orsay, France}}

\author{E. Frantzeskakis}
\email[e-mail: ]{emmanouil.frantzeskakis@universite-paris-saclay.fr}
\affiliation{\mbox{Universit\'{e} Paris-Saclay, CNRS, Institut des Sciences Mol\'{e}culaires d’Orsay, 91405, Orsay, France}}

\date{\today}
\begin{abstract}
Before assessing the suitability of a semiconductor for specific applications, the first question to ask is whether it possesses a direct or indirect band-gap. This distinction is fundamental, as the operation of devices such as light-emitting diodes, solar cells, and photodetectors is closely tied to the band-gap nature. Semiconductors that exhibit a band-gap crossover - indirect to direct or vice versa - offer enhanced versatility for optoelectronic applications. Prominent examples include transition metal dichalcogenides and the subject of this study, PbI$_2$. The nature of the band-gap, and its crossover, can only be directly determined in reciprocal space by tracking the valence-band maximum and conduction-band minimum. Here, we directly visualize the thickness-dependent crossover of PbI$_2$ from an indirect to a direct band-gap using angle-resolved photoemission spectroscopy. Our measurements reveal a shift of the valence-band maximum toward the Brillouin-zone center as the film thickness exceeds a monolayer. Supported by density functional theory calculations, our results show that this crossover is driven by interlayer interactions and the hybridization of iodine $p_z$ orbitals. These findings demonstrate the tunable electronic structure of PbI$_2$ and its potential for optoelectronic applications.

\end{abstract}
\maketitle
    
\section{Introduction}
\label{sec:intro}
The nature of the band-gap in a given semiconductor is crucial for its
applications: a photon with an energy of the order of 1~eV, 
close to band gaps of most semiconductors,
carries a momentum of the order of $5\times10^{-4}$~\AA$^{-1}$. This value
is negligible compared to the typical Brillouin-zone sizes of solids 
(of the order of $1$~\AA$^{-1}$) and to the 
separation, in reciprocal space,
between the valence band maximum (VBM) and the conduction band minimum (CBM) in indirect-gap semiconductors.
Thus, a direct band-gap favors optical absorption and radiative emission
(ideal for lasers and LEDs \cite{nakamura1994,ponce1997,holonyak1962,fadaly2020}), but also results in shorter electron-hole pair lifetimes. On the other hand, an indirect band-gap, where optical transitions between
the valence and conduction bands require the assistance of phonons, results in less efficient optical absorption (increased transparency), but also yields an increase in exciton
lifetimes through the inhibition of spontaneous radiative emission. Increased carrier lifetimes and the absence of spontaneous light emission open up applications in Si-based solar cells \cite{lin2023,wang2024}, mechanical sensors \cite{richter2005,munguia2008}
and low-optical-noise photonic circuits \cite{malik2021,aldaya2017}.

Two-dimensional layered semiconductors have attracted significant research interest as platforms for studying fundamental quantum phenomena \cite{Cha2023,Lin2014,Wang2020,Butler2020,Nakata2021}, but most importantly due to their applications in nanoelectronics and optoelectronics \cite{fiori2014,Xia2014, wang2012,Chhowalla2013,gouadria2024,prasad2023,cheng2014,ho2017}; an application potential that is closely related to the nature of their band-gap. Among the various 2D layered semiconductors, lead iodide (PbI$_2$) stands out with wide use in X-ray detectors \cite{SHAH1996266,SHAH2001140, mcgregor1997}, and as a precursor for organolead halide perovskites that offer considerable prospects in light-emitting diodes \cite{tan2014,guo2022}, photodetectors \cite{spina2015,hong2024} and solar cells \cite{Lee2012, Green2014}. A peculiar characteristic of PbI$_2$ is that its direct band-gap, the key feature behind its applications, becomes indirect when its thickness decreases to a single monolayer \cite{Shen2017,yagmurcukardes2018,zhong2017,wang2018,wang2024bis,toulouse2015}. This band-gap crossover in PbI$_2$ strongly affects its optoelectronic and vibrational properties \cite{Shen2017,yagmurcukardes2018,zhong2017}, while opening up a different spectrum of applications \cite{hoat2019,zhong2017,le2020,vo2020}. The thickness-dependent band-gap crossover in 2D layered semiconductors is, however, not exclusive to PbI$_2$. Transition metal dichalcogenides (TMDs), also, exhibit a change in the nature of their band-gaps in the monolayer form \cite{huo2014,lopez-sanchez2013,splendiani2010,mak2010,zhao2013}. The band-gap in TMDs follows, however, an opposite trend than in PbI$_2$, with a direct band-gap in monolayer TMD films and an indirect gap in their bulk form.

The thickness-dependent band-gap crossover in PbI$_2$ has been theoretically predicted by band structure calculations, which agree that for a monolayer film the VBM is located away from the center of the Brillouin zone, while it shifts to the latter when interlayer coupling comes into play, thereby altering the electronic structure in multilayer films \cite{Shen2017,yagmurcukardes2018,zhong2017,wang2018,wang2024bis,toulouse2015}. The mechanism for the band-gap crossover should therefore have a direct fingerprint in the experimental electronic structure of PbI$_2$. However, to this day, the experimental verification of the band-gap crossover only comes by indirect experimental means, such as photoluminescence and optical measurements that lack sensitivity in momentum space \cite{zhong2017,toulouse2015,sonny2025}.

Here, we present direct experimental evidence on the electronic structure of PbI$_2$ for the indirect-to-direct band-gap crossover as its thickness increases beyond a single monolayer. By means of angle-resolved photoemission spectroscopy (ARPES), we investigated the electronic structure of PbI$_2$ thin films deposited on graphene/6H-SiC(0001) \cite{pellecchi2014,benaziza2017}. Through a systematic study of films with different thicknesses, we were able to track the evolution of the experimental electronic structure and reveal the change in the nature of the band-gap. Our experimental results are in excellent agreement with density functional theory (DFT) calculations and pinpoint the exact location of the energy gap in reciprocal space.

\section{Methods}
\label{sec:meth}
\subsection{Sample growth}
\label{subsec:growth}

\begin{figure}[b]
    \centering
    \includegraphics[width=0.47\textwidth]{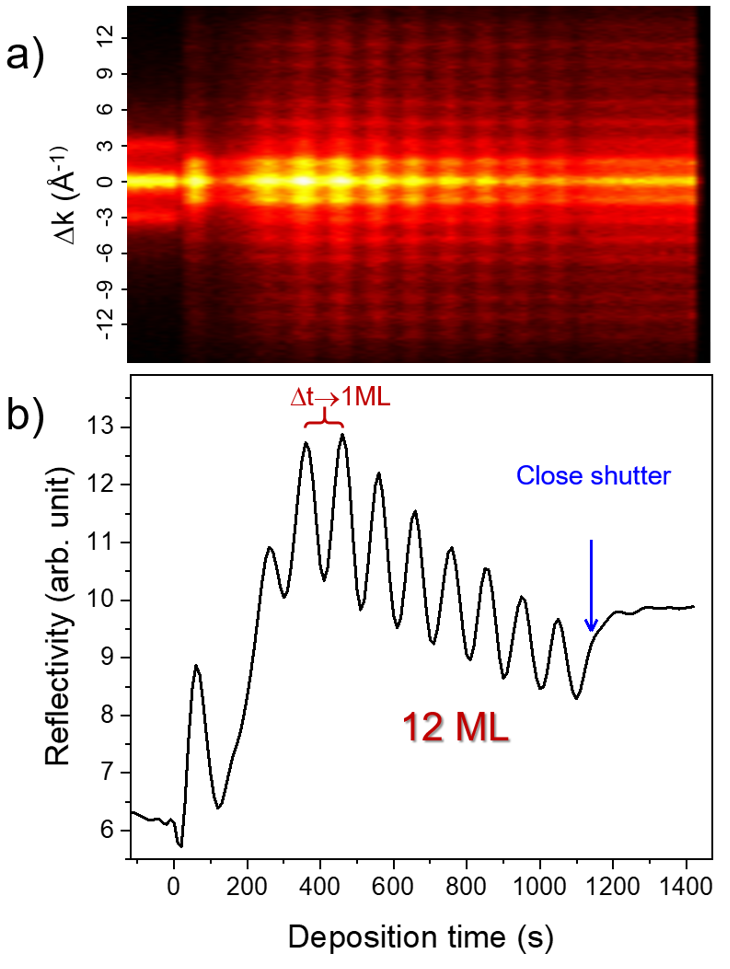}
    \caption{Growth dynamics of 12 layers of PbI$_2$ on graphene/SiC(0001), probed by a 600 eV He beam aligned along the graphene zigzag direction. (a) Evolution of the GIFAD pattern. (b) Reflectivity curve.}
    \label{fig:GIFAD}
\end{figure}

The sample growth was conducted in a homemade Molecular Beam Epitaxy chamber at a base pressure in the low $10^{-10}$ mbar range. Graphene substrates, grown on 6H-SiC(0001), were prepared by annealing cycles at $520\,^\circ$C until the observation of a well-contrasting GIFAD pattern. The deposition was carried out by heating the PbI$_2$ crucible at $300\,^\circ$C, while the substrate is at room temperature.

The growth dynamics was monitored in real time by Grazing Incidence Fast Atom Diffraction (GIFAD) \cite{khemliche2009,atkinson2014}. GIFAD relies on the coherent scattering of keV-energy He atoms at incidence angles typically lower than $1^\circ$ with respect to the surface plane. In contrast to X-ray or electron diffraction, GIFAD is a very soft technique that allows sensitive and damage-free characterization of the topmost layer only, with unambiguous information on film thickness. The latter is directly accessible through the intensity oscillations of the reflected atoms. Fig.~\ref{fig:GIFAD}(a) shows the time evolution of the GIFAD pattern during the growth of a 12-layer thick PbI$_2$ sample. The probe beam was parallel to the zigzag direction of graphene. The well-resolved structure along $\Delta k$ demonstrates a good relative alignment between the PbI$_2$ and substrate lattices \cite{sonny2025}, with armchair (PbI$_2$) $\parallel$ zigzag (graphene). The sharp contrast of the time-dependent oscillations, reproduced in Fig.~\ref{fig:GIFAD}(b), indicates a layer-by-layer growth mode, while the high intensity, relative to that measured on the bare substrate, reflects the high quality of the sample.

\subsection{Angle-resolved photoemission spectroscopy}
\label{subsec:ARPES}
Angle-resolved photoemission measurements were performed at the CASSIOPEE beamline of Synchrotron SOLEIL, France. The samples were transported in a UHV suitcase, directly after growth, with the pressure never exceeding \(5 \times 10^{-8}\)~mbar during the transfer. During the ARPES measurements, the pressure remained below \(3 \times 10^{-10}\)~mbar, and the sample temperature was stabilized at 16~K. Data were acquired using photon energies of 80~eV and 65~eV and linear horizontal polarization. The energy resolution was maintained at 30~meV.

\begin{figure*}
\includegraphics[width=\textwidth]{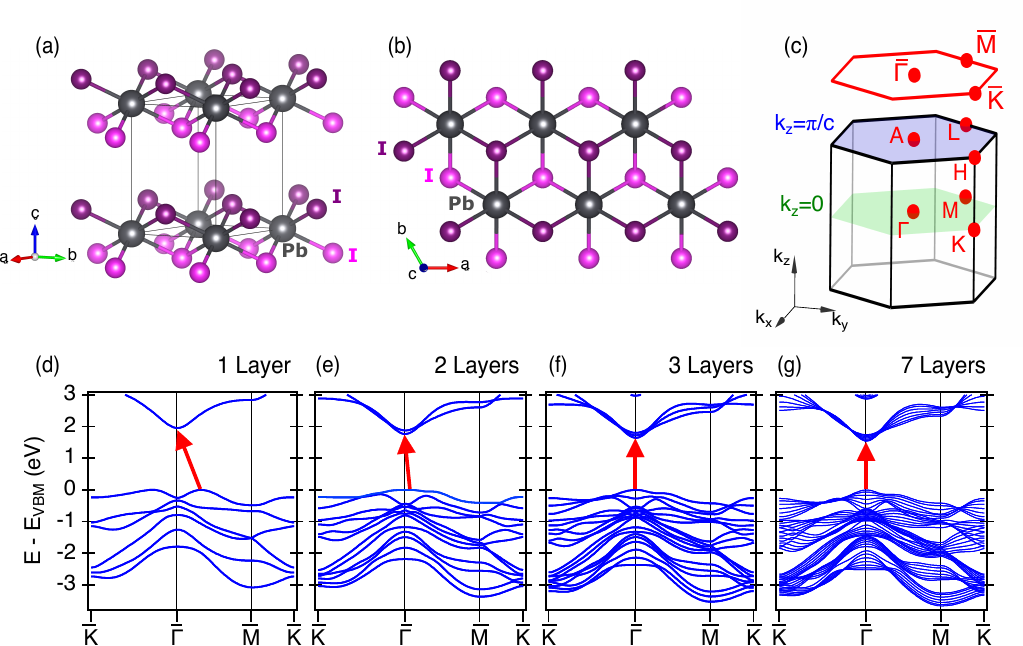}%
\caption{Structural and electronic properties of the 2H polytype of PbI$_2$.
(a) Atomic structure of PbI$_2$, showing a sandwich-like arrangement where a layer of Pb atoms (grey) is positioned between two layers of I atoms (purple and pink).  
(b) Top view of the monolayer atomic structure, revealing a hexagonal arrangement of iodine atoms surrounding each lead atom.  
(c) Bulk and surface Brillouin zone with the $k_z=0$ and $k_z=\pi/c$ planes highlighted in green and blue, respectively. (d)–(g) Theoretical band structure of PbI$_2$ obtained from GGA PBE calculations, for different number of layers. The calculations were performed for free-standing PbI$_2$. For multilayers, a slab approach has been used, thus resulting in a multitude of bands that evolve into a continuum in thicker films. Red arrows indicate the position of the minimum band-gap, with tilted (vertical) arrows denoting indirect (direct) band-gaps. The energy axis of each panel has been scaled with respect to its valence band maximum.}
\label{fig:structure}
\end{figure*}

\subsection{First-principles calculations}
\label{subsec:DFT}
First principles density functional theory (DFT) calculations were performed using the projected augmented-wave (PAW)~\cite{blochl.94} potentials implemented in the Vienna Ab initio Simulation Package (VASP)~\cite{kresse.hafner.93,kresse.furthmuller.96,kresse.joubert.99} code.
For the exchange-correlation energy, the generalized gradient approximation (GGA) in the Perdew--Burke--Ernzerhof (PBE) parametrization~\cite{perdew.burke.96} was used.
The energy cutoff for the plane-wave expansion was set to $400$~eV. 
Furthermore, the van der Waals correction was introduced within the DFT-D3 method with the Becke-Johnson damping function~\cite{grimme.ehrlich.11}. 
The bulk structure was optimized with a $6 \times 6 \times 4$ $k$-grid within the Monkhorst--Pack scheme~\cite{monkhorst.pack.76}. 
As a convergence criterion of the optimization loop, we took the energy change below $10^{-6}$~eV and $10^{-8}$~eV for the ionic and electronic degrees of freedom, respectively. The 
optimized structure was used to construct slabs with different number of PbI$_{2}$ layers.

\section{Results \& Discussion}
\label{sec:results}
\subsection{Crystal structure}
\label{subsec:struct}

The crystal structure of PbI$_2$ is schematically depicted in Figs.~\ref{fig:structure}(a) and \ref{fig:structure}(b). The PbI$_2$ monolayer adopts a hexagonal in-plane arrangement with Pb atoms sandwiched between two layers of I atoms. Each Pb atom is surrounded by six I atoms, forming a nearly octahedral environment. The bulk structure consists of stacked layers held together by weak van der Waals forces, forming the 2H polytype, the most stable phase of PbI$_2$ \cite{linPbI2SingleCrystal2019, minagawa1975, flahautCrystallization2H4H2006, beckmann2010, wangyang2016}, which is the structure of the multilayered films studied in this work (see subsection~\ref{subsubsec:ce}). We note that previous ARPES studies on bulk PbI$_2$ have focused on the 4H polytype, in which adjacent layers are arranged in a symmetric reflection with respect to the $c$-axis~\cite{Cha2023}. 
Bulk PbI$_2$ crystallizes in the $P\bar{3}m1$ space group (No.~164) with lattice parameters $a = b = 4.56$~\AA~and $c = 6.99$~\AA~\cite{Zhou2015}. 
The surface and bulk Brillouin zones (BZs) are presented in Fig.~\ref{fig:structure}(c), along with their high-symmetry points. 

\subsection{Theoretical electronic structure}
\label{subsec:thelec}

\begin{figure*}
\includegraphics[width=\textwidth]{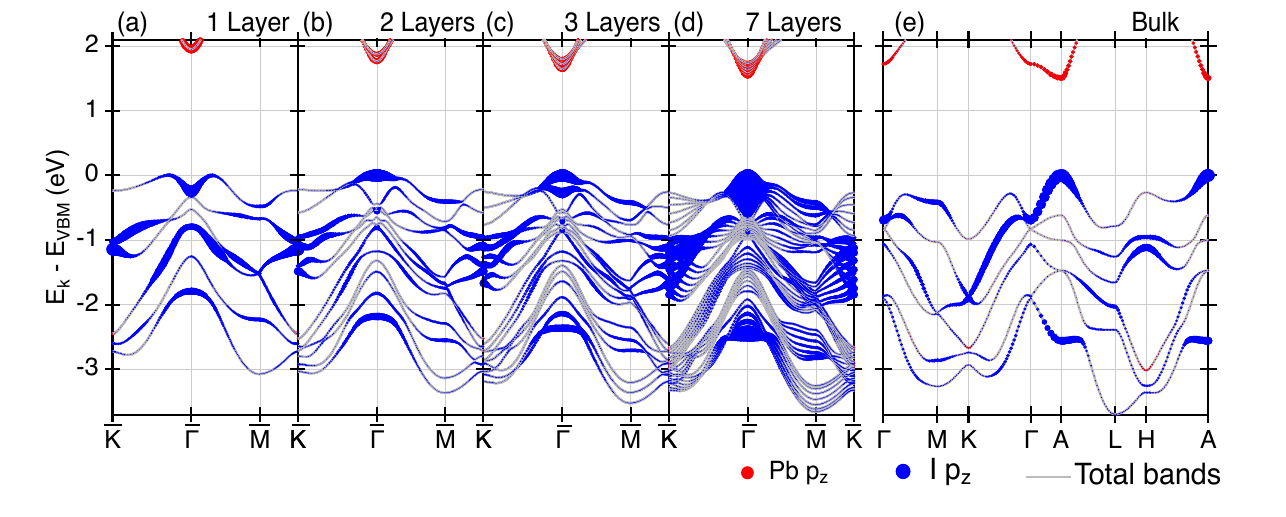}%
    \caption{
Orbital-resolved band structure for (a) one-, (b) two-, (c) three-, (d) seven-layer PbI$_{2}$ films, and (e) the bulk compound. 
Grey lines represent the full DFT-calculated bands, while red and blue points indicate the projection onto the $p_z$ orbitals of Pb and I atoms, respectively. The band structure of the thin films refers to the 2D surface-projected Brillouin zone, while the one of the bulk compound refers to the 3D bulk Brillouin zone [see panel (c) of Fig.~\ref{fig:structure}].
The origin of all energy axes has been set to the valence band maximum.
}
\label{fig:Sup_Orbitals}
\end{figure*}

We will now discuss the electronic structure of PbI$_{2}$ films
as predicted by our DFT calculations.
The most important change as a function of film thickness is the emergence of a direct band-gap, which has an immediate impact on the properties and technological applications of PbI$_2$ \cite{lan2017,sheng2015,ambardar2024b}. Fig.~\ref{fig:structure}(d)-(f) shows the evolution of the electronic structure with an increasing number of PbI$_2$ layers. For a single layer [panel (a)] the band-gap is indirect due to the fact that the VBM is not located at a high-symmetry point of the surface Brillouin zone (SBZ), but rather in-between $\bar{\Gamma}$ and $\bar{\text{M}}$. As the thickness increases [Figs.~\ref{fig:structure}(e) and~\ref{fig:structure}(g)], the VBM shifts towards $\bar{\Gamma}$ giving rise to a direct band-gap for films of three layers (3L) and beyond. The calculated gap size slightly varies as a function of thickness, ranging from $\sim$1.95 eV for a single layer, to $\sim$1.54 eV for 7L films. Since the location of the CBM is not affected by the increasing number of layers, we underline the key role of the VBM in establishing the nature of the band-gap: as the thickness increases beyond 1L, the emergence of electronic bands with a maximum at $\bar{\Gamma}$ determines the position of the VBM, which stays at $\bar{\Gamma}$ even after the development of the bulk continuum. 

The orbital-resolved band-structure is shown in Fig.~\ref{fig:Sup_Orbitals}. The gray lines represent the full DFT-calculated bands, while the different colors indicate projections onto the $p_z$ orbitals of the Pb (red) and I (blue) atoms. As thickness increases, the bands gradually approach the bulk-like dispersion. The conduction band is dominated by Pb $p_z$ states, while the top of the valence band in multilayer films has an I $p_z$ character. In fact, as the thickness increases, interactions between the I $p_z$ orbitals of neighboring layers result into the emergence of the electronic bands, that constitute the VBM at $\bar{\Gamma}$ in multilayer films \cite{yagmurcukardes2018}. 
The shift of the VBM to $\bar{\Gamma}$ is obvious when the film thickness is at least 3L [panels (f) and (g) in Fig.~\ref{fig:structure}], with the bilayer film [panel (e)] marking an intermediate stage between the monolayer behavior and the electronic structure of thicker multilayer films. The indirect-to-direct band-gap crossover of PbI$_2$ is in agreement with previous theoretical studies~\cite{yagmurcukardes2018,toulouse2015,zhong2017, wang2018}. It has also been used to explain the findings of experimental measurements with no direct view on the electronic structure of PbI$_2$ ~\cite{zhong2017,sonny2025}. However, a direct experimental verification of the indirect-to-direct band-gap crossover has never been performed and it will be the main focus of subsection~\ref{subsec:expelec}.

\subsection{Experimental Electronic structure} 
\label{subsec:expelec}
\subsubsection{Constant-energy surfaces}
\label{subsubsec:ce}

\begin{figure*}
\includegraphics[width=1\linewidth]{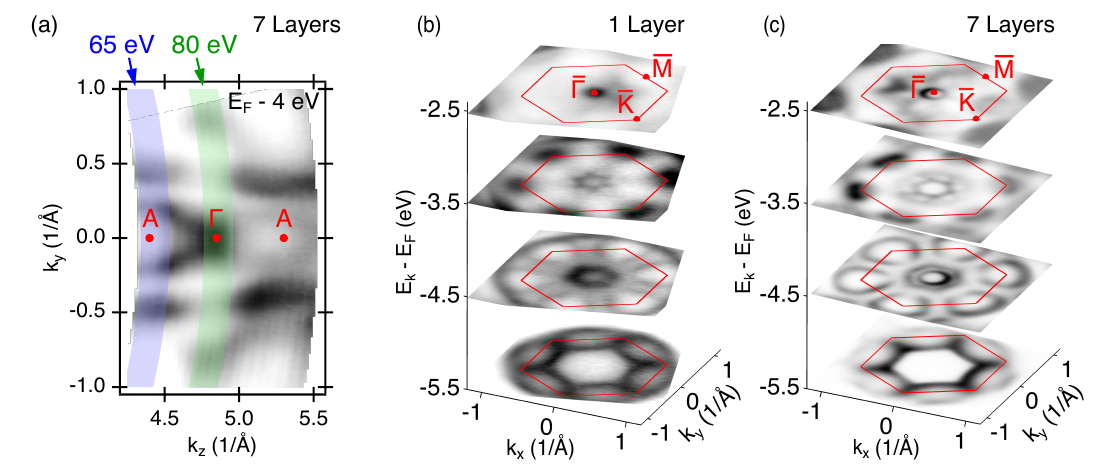}%
    \caption{(a) Out-of-plane constant energy map at a binding energy of $4$~eV, with green and blue arcs indicating constant photon energy contours. The thickness of the arcs approximates the $k_z$ broadening in ARPES and is equal to $\lambda^{-1}$, where $\lambda$ represents the photoelectron escape depth ~\cite{strocov2003}. The transformation to $k$-space was performed assuming an inner potential of $V_0 = 15$~eV. (b) and (c) Constant energy maps for monolayer and seven-layer PbI$_2$, respectively, measured at $k_z = 0$ (photon energy 80~eV), at a temperature of $16$~K, using linearly polarized light with horizontal polarization. Each panel presents four constant-energy contours for binding energies ranging from $5.5$~eV to $2.5$~eV, with a step of $1$~eV.}
\label{fig:const_maps}
\end{figure*}

The electronic structure of the 7L films being three-dimensional, one needs to identify the photon energies that allow us to probe the high-symmetry planes of the 3D Brillouin zone. To this end, we probed the constant energy contours in the plane perpendicular to the sample surface by means of photon-energy-dependent ARPES. The result is shown in Fig.~\ref{fig:const_maps}(a) at a binding energy for which the periodicity of the experimental features was most evident. We note that the out-of-plane periodicity of the electronic structure in Fig.~\ref{fig:const_maps}(a) ($2\pi/c=0.9$~Å$^{-1}$) is in excellent agreement with the out-of-plane constant of the 2H polytype of PbI$_2$ \cite{beckmann2010, minagawa1975, palosz1990,mitchell1959a}. Through a comparison to the theoretical electronic dispersion of bulk PbI$_2$ along the $\Gamma-\text{A}$ direction [see Fig. ~\ref{fig:bulk}(b) and Refs. \cite{yagmurcukardes2018, toulouse2015,Cha2023}], we identify the high-symmetry points corresponding to incident photon energies 65~eV and 80~eV, as, respectively, the A and $\Gamma$ points of the bulk Brillouin zone. We can therefore probe the electronic structure in the high-symmetry planes $k_z=0$ and $k_z=\pi/c$ [see  Fig.~\ref{fig:structure}(c)] using, respectively, 80~eV and 65~eV photons. Although the electronic structure of the 1L films is, as expected, 2D (see Appendix A), when we compare it to the 7L films, we used identical photon energies in order to account for the effect of the photoemission matrix elements that are highly dependent on the experimental conditions \cite{moser2017}.
 
The in-plane constant energy surfaces observed in Fig.~\ref{fig:const_maps}(b) and ~\ref{fig:const_maps}(c) exhibit an obvious hexagonal symmetry that matches the periodicity of the surface-projected Brillouin zone (SBZ). The energy planes correspond to the same binding energy in both panels varying from $5.5$~eV (bottom) to $2.5$~eV (top). We note that, due to the insulating character of PbI$_2$, the Fermi energy was determined by the metallic states of graphene originating from the substrate (see Appendix B). Although the overall shape of the constant-energy contours does not change substantially between the 1L and 7L films, the observed differences are due to changes of the electronic structure with increasing thickness. Such changes comprise the evolution of well-separated states into a band continuum as predicted by theory [Fig.~\ref{fig:structure}(d)-(g)], slight modifications of the fine details in the energy-momentum dispersion and the relative energy shift of some electron bands (see subsection~\ref{subsubsec:comp} for more details).

\subsubsection{Experimental fingerprints of the band-gap crossover}
\label{subsubsec:crossover}

\begin{figure}[h!]
\includegraphics[width=\linewidth]{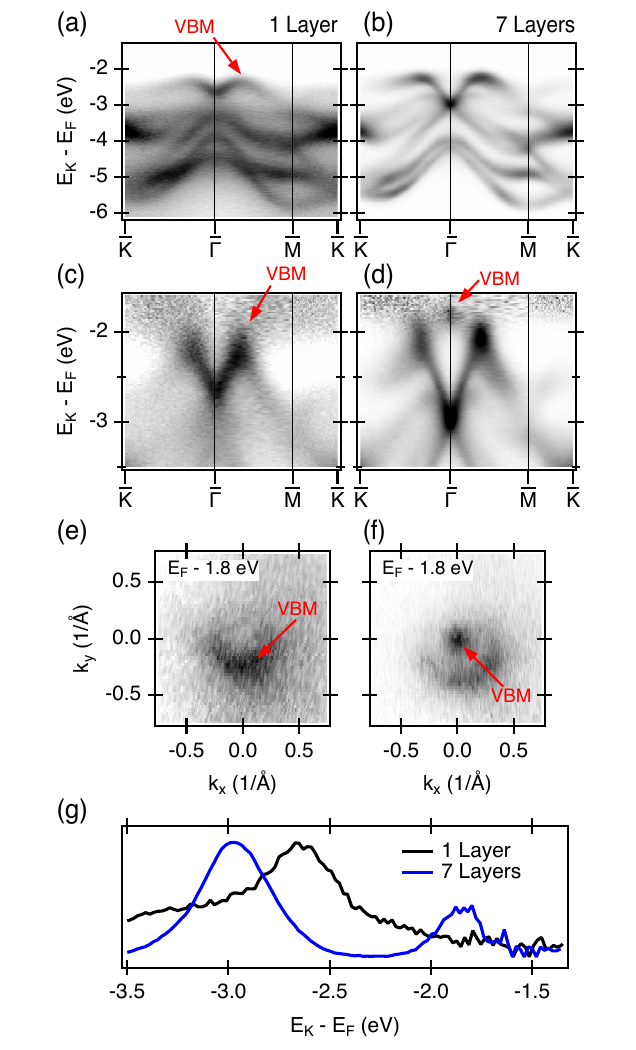}%
    \caption{Electronic structure of 1 layer (1L) and 7 layers (7L) PbI$_2$ films. 
    (a) and (b) Band structure of, respectively, 1L and 7L films along the $\bar{\text{K}}$-$\bar{\Gamma}$-$\bar{\text{M}}$-$\bar{\text{K}}$ path. (c) and (d) Band structure of, respectively, 1L and 7L films with maximally enhanced contrast for each momentum distribution curve (MDC). The contrast enhancement was achieved by normalizing each MDC to its integrated intensity. The tail of the red arrows indicates the location of the VBM for both films. (e) and (f) Constant energy maps corresponding to a binding energy of $1.8$~eV (i.e. at the valence band maximum) for 1L and 7L films, respectively, under identical measurement conditions. (g) Energy distribution curves (EDCs) corresponding to panels (c) and (d), obtained by integrating over a momentum window of 0.1~Å$^{-1}$ around $\bar{\Gamma}$. The blue (black) curve corresponds to 7L (1L) films. Measurements have been performed at $16$K using photons of $80$ eV and linear horizontal polarization.}
\label{fig:VBM}
\end{figure}

Figure~\ref{fig:VBM} shows the experimental electronic structure of 1L (left) and 7L (right) PbI$_2$ thin films. The valence band consists of a multitude of bands that span the approximate energy range from $-6$~eV to $-2$~eV [Figs.~\ref{fig:VBM}(a) and \ref{fig:VBM}(b)]. The overall shape of the energy-momentum dispersion for the 1L and 7L films is quite similar, but there are certain differences which are captured by our DFT calculations. For example, the monolayer film shows an energy gap at $\bar{\Gamma}$ just below the topmost (``wavy-shaped'') band, while the thicker film exhibits a high-intensity feature indicative of a band crossing. Such a band crossing can be seen in the theoretical band structure in Fig.~\ref{fig:structure}(g) where the two topmost occupied band ``ribbons'' meet each other at approximately 1~eV below the VBM. Another example is the shape evolution of the lowermost experimental state of the valence band, which becomes more dispersing with increasing thickness. This evolution is theoretically captured in Fig.~\ref{fig:structure}(d)-(g), where the lowermost state of the valence band is rather flat around $\bar{\Gamma}$ for monolayer films, while it becomes more dispersing as it evolves into a band continuum. More details into the comparison of the electronic structure of 1L and 7L films are presented in subsection~\ref{subsubsec:comp}.

The most crucial element in comparing the experimental electronic structure of 1L and 7L PbI$_2$ films is the $k$-space location of the VBM. The band dispersions along the $\bar{\text{K}}$–$\bar{\Gamma}$–$\bar{\text{M}}$–$\bar{\text{K}}$ path, shown in Figs.~\ref{fig:VBM}(a) and ~\ref{fig:VBM}(b), reveal a similar overall shape for both thicknesses. However, due to the relatively low intensity of the topmost valence states, the precise position of the VBM is not directly obvious from these raw intensity plots. To better visualize weak spectral features near the valence band edge, we applied a contrast enhancement method (see caption). The resulting enhanced-contrast band maps, shown in Fig.~\ref{fig:VBM}(c) and Fig.~\ref{fig:VBM}(d), provide a clearer view of the VBM. In the monolayer film [Fig.~\ref{fig:VBM}(c)], the VBM is located at the top of a wavy-shaped band, clearly off-centered from $\bar{\Gamma}$. The $k$-space separation between the VBM in monolayer films and the center of the Brillouin zone is clear and equal to $0.22$~\AA$^{-1}$. 
The corresponding energy difference between the topmost band at  $\bar{\Gamma}$ and the real VBM is about 0.3 eV. 
On the contrary, the 7L film [Fig.~\ref{fig:VBM}(d)] reveals a faint additional state with a maximum right at $\bar{\Gamma}$; a state that was absent in the 1L case. This state, derived from I $p_z$ orbitals and discussed in the context of Figs.~\ref{fig:structure} and~\ref{fig:Sup_Orbitals}, constitutes the actual VBM in the 7L film and marks the crossover from an indirect to a direct band-gap. As it will be explained in subsections~\ref{subsubsec:comp} and \ref{subsubsec:loc}, the detailed dispersion of the state constituting the VBM for 7L is better resolved at different experimental conditions than those of Fig.~\ref{fig:VBM}(d). The corresponding data is shown in panel (c) of Fig.~\ref{fig:Sup_1L7L}. In the context of Fig.~\ref{fig:VBM}, we chose to keep identical experimental conditions for the comparison of the 1L and 7L data to avoid complications related to the matrix elements effects of the photoemission process \cite{moser2017}. 

The band-gap crossover is also evident when inspecting the constant energy contours at the VBM, shown in Fig.~\ref{fig:VBM}(e) and Fig.~\ref{fig:VBM}(f) for 1L and 7L, respectively. In the monolayer film, the contour forms a ring-like shape with no central feature, while in the 7L case, an additional dot-like intensity feature appears at $\bar{\Gamma}$, a direct fingerprint of the VBM moving to the center of the Brillouin zone. To further support this observation, Fig.~\ref{fig:VBM}(g) displays energy distribution curves (EDCs) extracted from the contrast-enhanced maps in Fig.~\ref{fig:VBM}(e) and Fig.~\ref{fig:VBM}(f). The EDC for the 7L film (blue) shows a clear spectral feature at the VBM, consistent with the presence of the central state, while the EDC for the 1L film (black) does not exhibit this feature.

\subsubsection{Band-gap location in 3D reciprocal space}
\label{subsubsec:loc}

Figure~\ref{fig:VBM} establishes the in-plane $k$-space location of the VBM in 1L and 7L films. However, the 7L films have a three-dimensional electronic structure with an out-of-plane periodicity that corresponds to bulk PbI$_2$ crystals of the 2H polytype (see subsection~\ref{subsubsec:ce} for more details). It would be therefore instructive to pinpoint the experimental $k_z$ location of the VBM in the 7L films and compare it to theoretical predictions for the bulk. 

\begin{figure}
\includegraphics[width=\linewidth]{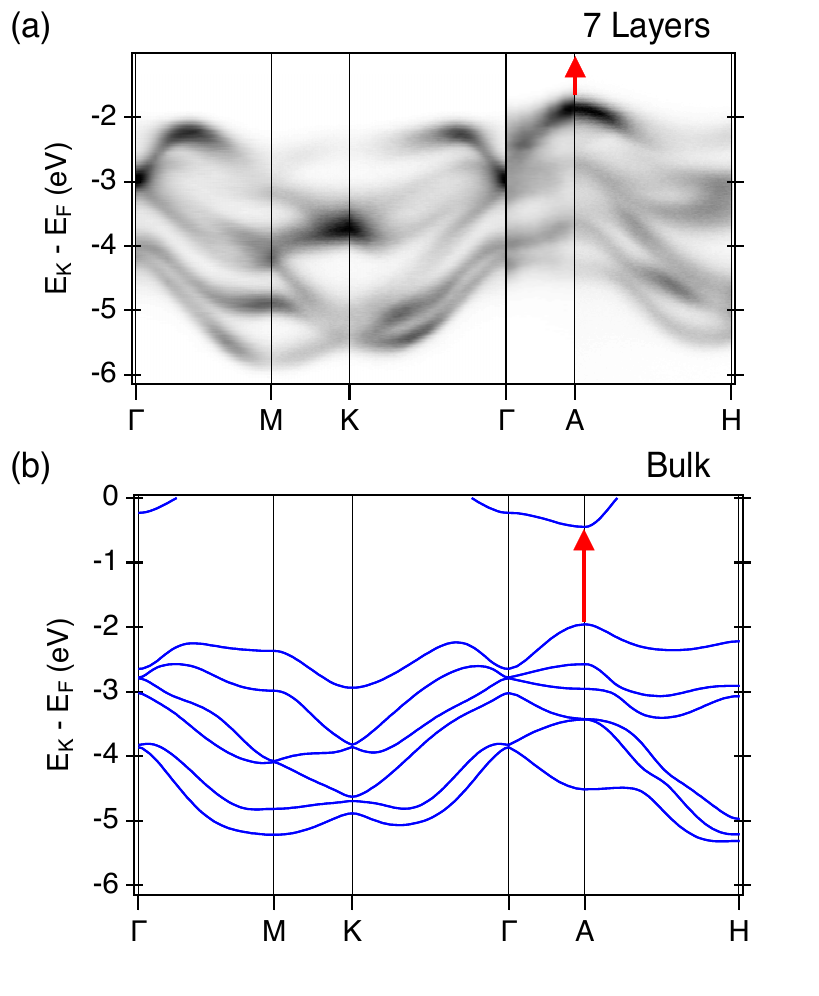}%
    \caption{Bulk band structure of PbI$_2$. (a) ARPES spectra along the $\Gamma$-$\text{M}$-$\text{K}$-$\text{G}$-$\text{A}$-$\text{H}$ path. The in-plane segments corresponding to $k_z = 0$ and $k_z = \pi/c$ were measured with photon energies of 80~eV and 65~eV, respectively. The out-of-plane $\Gamma$-$\text{A}$ segment was measured by systematically varying the photon energy with a 1~eV step. Data were acquired at $T = 16$~K using photons with linear horizontal polarization.  (b) Theoretical band structure for bulk PbI$_2$ obtained from GGA PBE calculations along the same high symmetry directions. Red arrows indicate the position of the minimum band-gap. The energy scale in (b) has been rigidly shifted in order to match the experimental energies in (a), and thus enable a direct comparison of theory and experiment. 
}
\label{fig:bulk}
\end{figure}

Figure~\ref{fig:bulk}(a) shows the three-dimensional experimental electronic structure of PbI$_2$ 7L films, including the out-of-plane directions. Data in the $k_z=0$ and $k_z=\pi/c$ planes, schematically shown in Fig.~\ref{fig:structure}(c), have been respectively acquired with 80~eV and 65~eV incident photons, following the identification of the bulk high-symmetry points by ARPES (see subsection~\ref{subsubsec:ce}). All experimental bands can find their counterparts in our bulk DFT calculations [Fig. \ref{fig:bulk}(b)], thereby showing an excellent agreement between theory and experiment. There is a clear dispersion along the out-of-plane $\Gamma$-$\text{A}$ path of the bulk Brillouin zone, characteristic of the three-dimensional character of the electronic structure. 
Most importantly, the VBM is located at the A point in line with the theoretical band diagram in Fig.~\ref{fig:bulk}(b) and with past literature~\cite{yagmurcukardes2018, toulouse2015, Cha2023}. We deduce an energy value of $\sim$1.8~eV for the VBM of the 7L films, in comparison to $\sim$2.3~eV for the VBM of the 1L films [Fig.~\ref{fig:VBM}(a)]. This observation amounts to a reduction in the gap size by 0.5 eV, since the energy position of the conduction band minimum is thickness-independent~\cite{yagmurcukardes2018}. The experimentally-observed decrease of the gap size is in good agreement with the predictions of our DFT calculations that infer an indirect band-gap of $\sim$1.95~eV for 1L films and a direct band-gap of $\sim$1.54~eV for 7L films (Fig.~\ref{fig:structure}). It should be however noted that the theoretical band gap
size obtained is underestimated with respect to the experimental value due to 
limitations of calculations within the GGA PBE framework ~\cite{jonesDensityFunctionalTheory2015, grumetQuasiparticleApproximationFully2018, leeFirstprinciplesApproachPseudohybrid2020}.
However, the main features of the band structure remain qualitatively unchanged. In this study, our theoretical calculation did not aim to determine the exact band-gap values but rather to probe the evolution of the electronic structure with increasing thickness. The two principal findings, i.e., the indirect-to-direct band-gap crossover [(Fig.~\ref{fig:structure})(d)-(g)] and the exact $k$-space location of the VBM in thick PbI$_2$ films [Fig.~\ref{fig:bulk}(b)] are both corroborated by our experimental results.

Given that the VBM is predicted [and experimentally probed in Fig.~\ref{fig:bulk}(a)] to lie in the $k_z=\pi/c$ plane (i.e.,~$h\nu=65$ eV), one might wonder why it can be still observed in Fig.~\ref{fig:VBM}(f), even if the latter is measured in the $k_z=0$ plane (i.e., $h\nu=80$ eV). This can be explained by the limitations of the ARPES technique. Namely, ARPES has relatively poor resolving power along the $k_z$~direction, which leads to the intermixing of signals from adjacent $k_z$ values~\cite{strocov2003}. As a result, the observed spectrum is a projection of a part of the 3D electronic structure onto the 2D plane of the SBZ. Consequently, even when measurements are performed at a photon energy corresponding to $k_z = 0$, there can still be a weak contribution from bands associated with other $k_z$ values. Thus, the faint feature observed at the top of the valence band in Fig.~\ref{fig:VBM}(f) can be attributed to the real VBM lying at $k_z=\pi/c$ that contributes a weak photoemission signal even at $k_z=0$.

\subsubsection{Band structure comparison of 1L and 7L PbI$_{2}$}
\label{subsubsec:comp}

\begin{figure*}
\includegraphics[width=\linewidth]{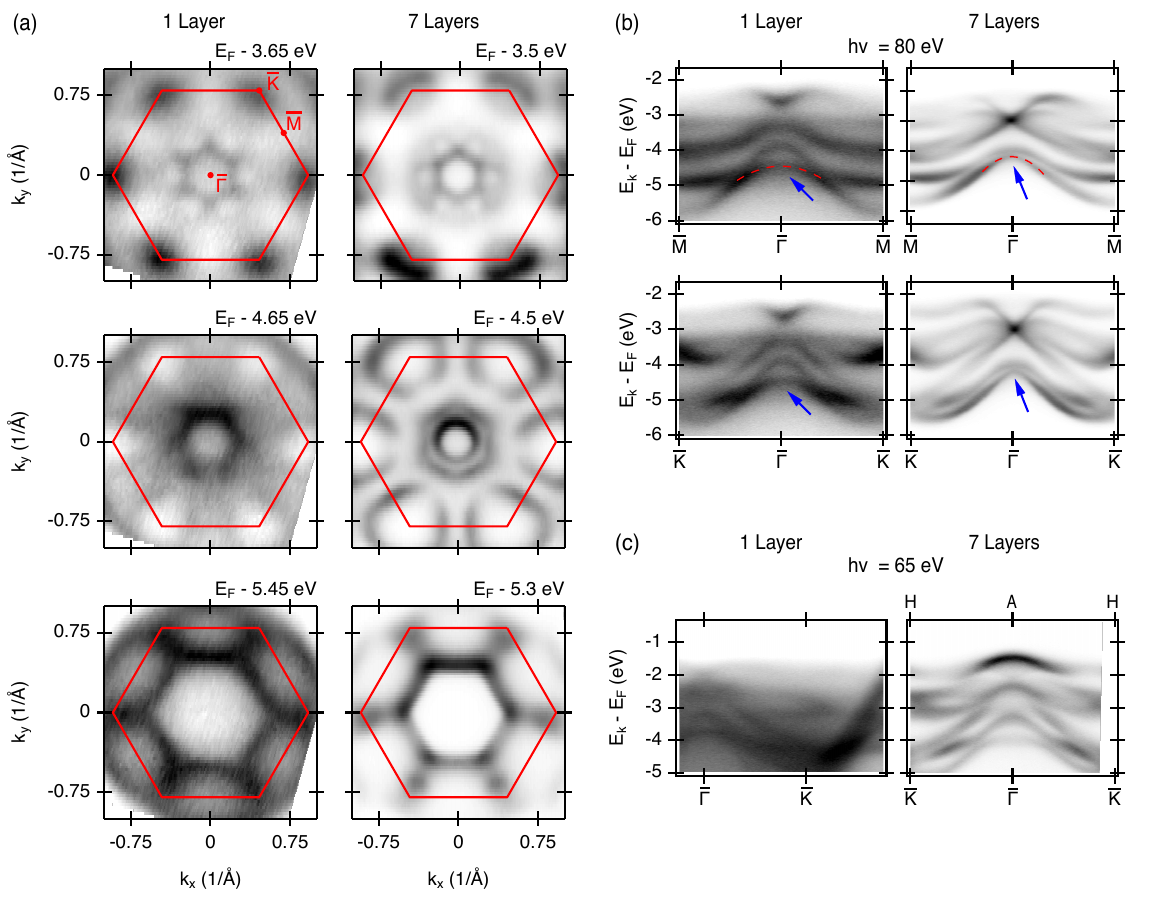}%
    \caption{Comparison of the electronic structure in 1L and 7L PbI$_{2}$ films. (a) Representative constant energy maps for 1L (left) and 7L (right) PbI$_{2}$, binding energies are indicated in the figure, red solid lines denote the borders of the surface Brillouin zone. (b) Comparison of the band structure along $\bar{\text{M}}-\bar{\Gamma}-\bar{\text{M}}$ (top) and $\bar{\text{K}}-\bar{\Gamma}-\bar{\text{K}}$ (bottom). Blue arrows indicate a state becoming more dispersive with increasing thickness. The red dashed lines are parabolic fits around $\bar{\Gamma}$. (c) Comparison of the band structure along $\bar{\text{K}}-\bar{\Gamma}-\bar{\text{K}}$ under experimental conditions that favor the observation of the VBM in 7L films. All spectra were measured at 16~K using photons of linear horizontal polarization and an energy of either 80 eV [panels (a), (b)] or 65 eV [panel (c)].}
\label{fig:Sup_1L7L}
\end{figure*}

We will now present further details on the electronic structure comparison between 1L and 7L PbI$_2$ films to extend the findings of the previous subsections and elucidate the valence band of the thicker films under the most favorable experimental conditions.

The evolution of the electronic structure of PbI$_2$ with increasing thickness is presented in Fig.~\ref{fig:Sup_1L7L}. Representative constant energy maps for the single-layer (1L, left panel) and seven-layer (7L, right panel) PbI$_2$ films are shown in Fig.~\ref{fig:Sup_1L7L}(a). In order to obtain similar features in the constant energy maps of the two films, a phenomenological energy shift of approximately 150~meV needs to be applied signaling charge transfer as thickness increases. Although the general features of the constant-energy contours are similar after the energy shift, remaining differences indicate that increasing thickness does not induce a simple rigid band shift, but also leads to a slight deformation of the electronic structure.

As briefly explained in subsection~\ref{subsubsec:crossover}, the comparison of the band structures [Fig.~\ref{fig:Sup_1L7L}(b)] reveals two main differences. First of all, for the monolayer sample, a clear energy gap is observed between the top electron-like band and the top hole-like band at the $\bar{\Gamma}$ point. On the contrary, for the 7L film, the corresponding bands cross each other, leading to an increase in spectral intensity at their crossing point. Secondly, the evolution of the lowermost hole-like state at $\bar{\Gamma}$ is clearly visible [dashed guides-to-the-eye and blue arrows in Fig.~\ref{fig:Sup_1L7L}(b)]. This state becomes more dispersing as the thickness increases, with its effective mass changing from -0.4$m_e$ (1L) to -1.1$m_e$ (7L).

In the context of Fig.~\ref{fig:VBM}, we compared the top of the valence band in 1L and 7L PbI$_2$ films under measurement conditions that favor the experimental observation of the topmost (i.e. ``wavy-shaped'') band which constitutes the VBM for the monolayer [Fig.\ref{fig:VBM} and Fig.~\ref{fig:Sup_1L7L}(b)]. Later, in the context of Fig.~\ref{fig:bulk}, it became nevertheless clear that these experimental conditions do not stay favorable for the topmost band of the 7L films due to its $k_z$ dispersion. In panel (c) of Fig. ~\ref{fig:Sup_1L7L}, we repeat this comparison under measurement conditions that now favor the observation of the VBM in 7L films (i.e. $h\nu=65 eV$, which corresponds to the $k_z=\pi/c$ plane). The 7L panel reveals a hole-like dispersion around the VBM that is in good qualitative agreement with the theoretical dispersions shown in Fig. ~\ref{fig:bulk}(b) (dispersion along $\Gamma-$A) and in Fig.~\ref{fig:structure} (g) (top part of the last occupied band ``ribbon" along $\bar{\Gamma}-\bar{\text{K}}$). On the other hand, we note that such experimental conditions are unfavorable for the aforementioned ``wavy-shaped'' band, and strongly enhance the electronic states of the graphene substrate (see increased intensity at the rightmost part in the 1L panel).

\section{Conclusions}
\label{sec:concl}

Our study presents a direct experimental demonstration of the band-gap crossover in lead iodide (PbI$_2$), a layered semiconductor that is receiving growing attention for optoelectronic applications. We employed angle-resolved photoemission spectroscopy (ARPES), the only experimental technique capable of directly probing the nature of the band-gap and its crossover where these notions are fundamentally defined, i.e. in reciprocal space. We investigated the evolution of the electronic structure in PbI$_2$ as a function of film thickness with our ARPES measurements revealing a pronounced shift of the valence band maximum toward the center of the Brillouin zone in films thicker than a monolayer. This provides clear experimental evidence for the crossover from an indirect to a direct bandgap. Although the indirect-to-direct band-gap crossover in PbI$_2$ has previously been predicted by band structure calculations~\cite{yagmurcukardes2018,toulouse2015,zhong2017, wang2018} and its consequences observed using techniques such as photoluminescence \cite{zhong2017}, our study presents the first direct experimental observation of the emergence of the direct band-gap from its indirect counterpart.

Apart from the band-gap crossover, our ARPES results experimentally demonstrate the expected dimensionality transition in the electronic structure of PbI$_2$ thin films with varying thickness. Monolayer PbI$_2$ exhibits a predominantly two-dimensional electronic character, whereas multilayer films show significant out-of-plane dispersion [Fig. \ref{fig:bulk}(a)], indicating the emergence of three-dimensional electronic behavior due to enhanced interlayer coupling. The electronic structure of 7-layer (7L) films closely matches that of the bulk, enabling a precise experimental identification of the valence band maximum in the three-dimensional reciprocal space. The experimentally determined band structures across different thicknesses show excellent agreement with our density functional theory (DFT) calculations, which predict that the band-gap crossover is driven by interlayer interactions and orbital hybridization, particularly involving iodine $p_z$ orbitals. This strong correspondence between experiment and theory serves as a robust validation of the experimental trends related to the electronic structure dimensionality and -most importantly- the thickness-dependent band-gap crossover in PbI$_2$ films.

From the perspective of electronic gap engineering, our findings offer crucial insight into the tunability of the band structure in PbI$_2$ through thickness control. This thickness-dependent modulation of the band-gap nature highlights PbI$_2$ as a promising candidate for next-generation optoelectronic and photonic devices, where tailored band structures are essential for optimizing device performance.

\begin{acknowledgments}
The authors acknowledge SOLEIL for the provision of synchrotron radiation under the proposal No. 20240208. We thank F. Bertran for his support during our experiments at the CASSIOPEE beamline. This work was supported by public grants from the French National Research Agency (ANR), projects SUPERNICKEL (No. ANR-21-CE30-0041-05), MICROVAN (No. ANR-24-CE30-3213-01) and FastNano (No. ANR-22-PEXD-0006), and by the CNRS International Research Project EXCELSIOR. Schematic drawings of the crystal structure were made using VESTA~\cite{Mommadb5098}.
\end{acknowledgments}

\renewcommand{\thefigure}{A\arabic{figure}} 
\setcounter{figure}{0} 

\section*{APPENDIX A: 2D character of the electronic structure of 1L {\boldmath$\mathrm{PbI_2}$}}
\label{sec:appa}

\begin{figure}
    \centering
    \includegraphics[width=1\linewidth]{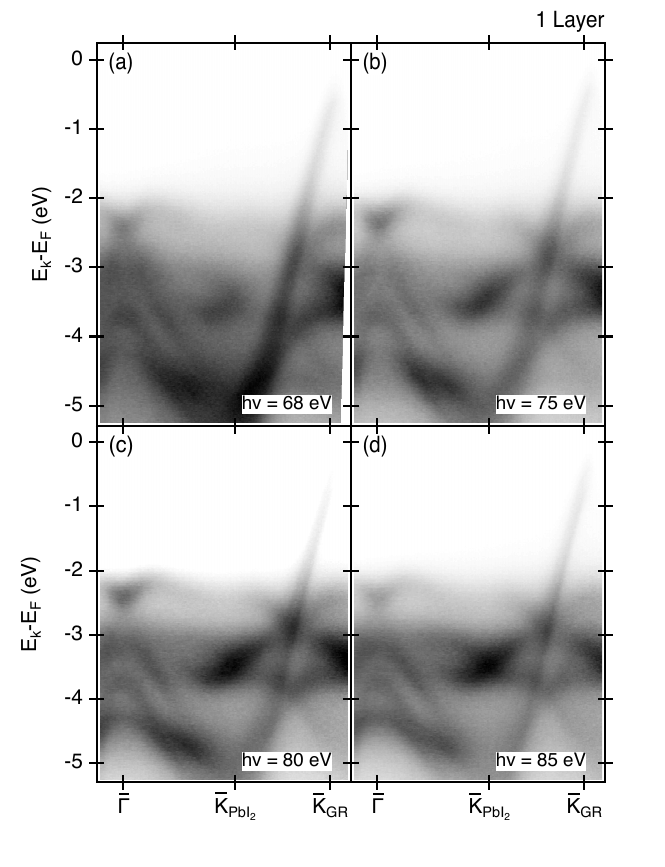}
    \caption{ARPES spectra of PbI$_2$ on graphene/SiC(0001) as a function of the photon energy: (a) $h\nu = 68$~eV, (b)~75~eV, (c)~80~eV, (d)~85~eV. All spectra were measured with linear horizontal polarization at $16$K.}
    \label{fig:1L_kz}
\end{figure}

In order to investigate the dimensionality of the electronic states in monolayer PbI$_2$, we have performed ARPES measurements over a broad photon energy range. Representative examples of the band structure recorded at four different photon energies, $h\nu = 68$~eV, 75~eV, 80~eV, and 85~eV, are shown in Fig.~\ref{fig:1L_kz}. The band dispersion remains essentially unchanged across the entire energy range, indicating the absence of measurable $k_z$ dependence. This behavior is consistent with expectations for a van der Waals layered material, where the electronic states are confined to the two-dimensional plane due to weak interlayer interactions. The lack of noticeable energy shifts or changes in the band structure with varying photon energy is in line with the expected two-dimensional character of the electronic states in monolayer PbI$_2$.

\begin{figure*}
\includegraphics[width=\textwidth]{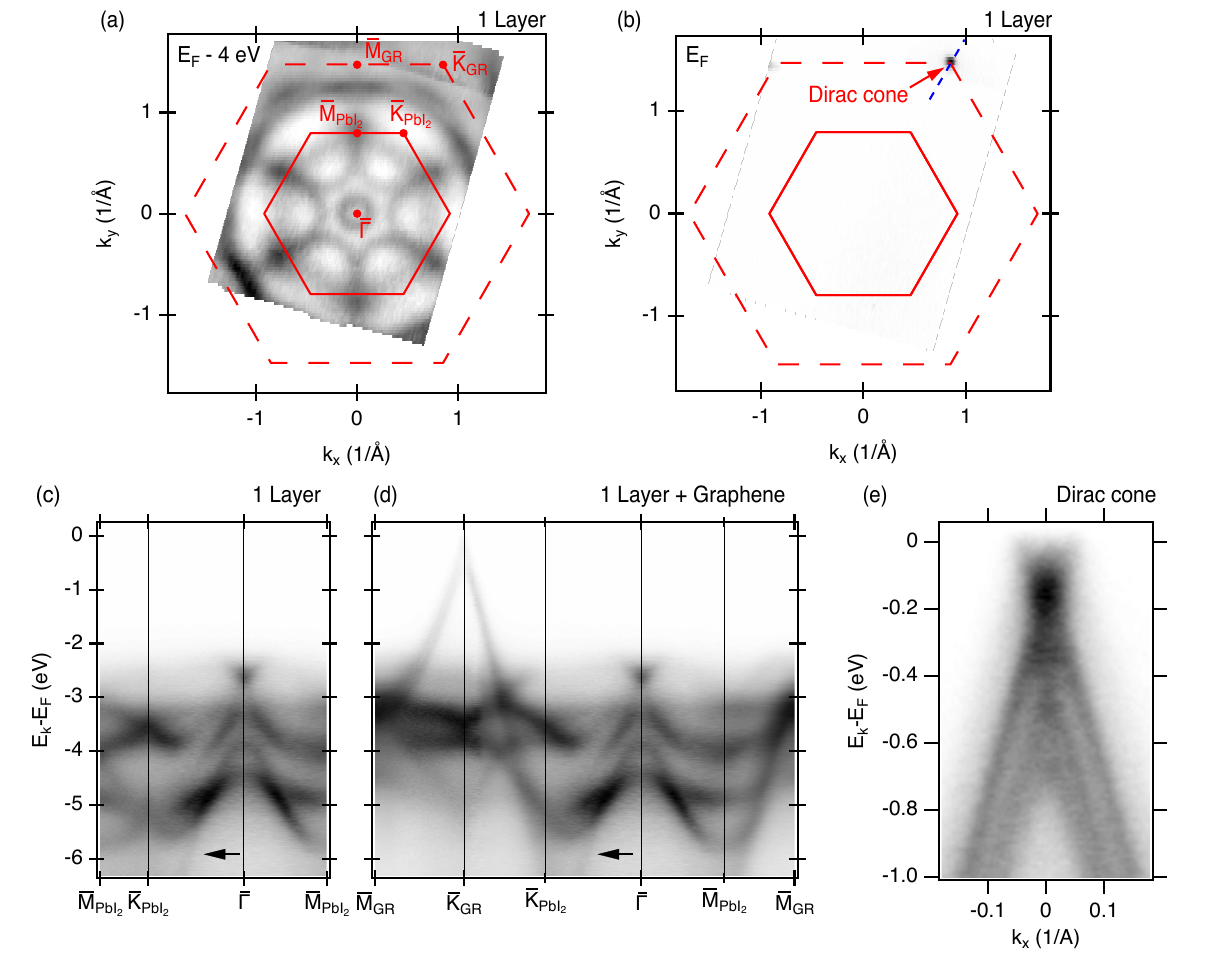}%
    \caption{Electronic structure of monolayer PbI$_{2}$ on graphene/SiC(0001). (a) Constant energy map at a binding energy of $4$~eV. (b) Fermi surface map. Red dashed and solid lines indicate the borders of the surface Brillouin zones of graphene and PbI$_{2}$, respectively. The position of the scan shown in (e) is marked by the blue dashed line. (c) Band structure of PbI$_2$ along the $k$-space path $\bar{\text{M}}$–$\bar{\text{K}}$–$\bar{\Gamma}$–$\bar{\text{M}}$ corresponding to its surface Brillouin zone. (d) Full band structure  showing contributions from both PbI$_2$ and the graphene substrate. (e) Dirac cone originating from the graphene substrate. Black arrows in (c) and (d) indicate the weakly visible band attributed to the graphene substrate. All spectra were measured at $16$K using photons of $80$ eV and linear horizontal polarization.}
\label{fig:Graphene}
\end{figure*}

\section*{APPENDIX B: Complete band structure of 1L  {\boldmath$\mathrm{PbI_2}$} and graphene}
\label{sec:appb}

Since the probing depth of photoemission in the UV range exceeds the thickness of a single PbI$_{2}$ layer, the ARPES spectra on monolayer PbI$_2$ include the spectral fingerprints of the graphene/SiC(0001) substrate. The complete electronic structure of a single-layer PbI$_2$ film grown on the graphene/SiC(0001) substrate is shown in Fig.~\ref{fig:Graphene}. The dashed and solid red lines denote, respectively, the borders of the graphene and PbI$_2$ surface Brillouin zones. A representative constant energy map is presented in Fig.~\ref{fig:Graphene}(a). The bands originating from PbI$_2$ form a star-like pattern centered at the $\bar{\Gamma}$ point, surrounded by a nearly circular feature arising from graphene. In the vicinity of the Fermi level [Fig.~\ref{fig:Graphene}(b)], the PbI$_2$-derived bands are absent, but one can readily observe an intense dot-like feature at the $\bar{\text{K}}$ high-symmetry point of graphene, originating by the so-called Dirac cone \cite{bostwick2007, castroneto2009, novoselov2005,ohta2006b,zhou2007}. The Dirac cone being a metallic state, was used to determine the Fermi energy for all of the spectra presented in the article. 

Figures~\ref{fig:Graphene}(c) and \ref{fig:Graphene}(d) illustrate the band structures of the PbI$_2$ layer and graphene along the $\bar{\text{M}}$–$\bar{\text{K}}$–$\bar{\Gamma}$–$\bar{\text{M}}$ path in their corresponding surface Brillouin zones. In the energy range shown in pnales (c) and (d), the graphene bands are primarily found around the $\bar{\text{K}}$ point, and have minimal overlap with the PbI$_2$ features. The only graphene-derived band, distinguishable within the PbI$_2$ region is a faint state near the PbI$_2$ $\bar{\text{K}}$ point, indicated by the black arrows.

The most prominent feature of the graphene electronic structure is the Dirac cone located at the $\bar{\text{K}}$ point. A detailed scan along the $\bar{\Gamma}-\bar{\text{M}}-\bar{\Gamma}$ direction is shown in Fig.~\ref{fig:Graphene}(e). The Dirac point is shifted below the Fermi level, a consequence of electron transfer from the SiC substrate to the graphene layers \cite{razado-colambo2018,coletti2010}.  Additionally, the characteristic double-cone shape arises from the bilayer nature of the graphene substrate with AB stacking \cite{silva2020,jin2021}.


\bibliography{PbI2}%






\end{document}